# Discrete breathers in an array of self-excited oscillators: exact solutions and stability.

I.B. Shiroky and O.V. Gendelman

Faculty of Mechanical Engineering, Technion – Israel Institute of Technology

**Abstract**

Dynamics of array of coupled self-excited oscillators is considered. Model of Franklin bell is adopted as a mechanism for the self-excitation. The model allows derivation of exact analytic solutions for discrete breathers (DBs), and extensive exploration of their stability in the space of parameters. The DB solutions exist for all frequencies in the attenuation zone, but lose stability via Neimark-Sacker bifurcation in the vicinity of the boundary of propagation zone. Besides the well-known DBs with exponential localization, the considered system possesses additional and novel type of solutions – discrete breathers with main frequency in the propagation zone of the chain. The amplitude of oscillations in this solution is maximal at the localization site and then exponentially approaches constant value at infinity. We also derive these solutions in closed analytic form. They are stable in a narrow region of system parameters bounded by Neimark-Sacker and pitchfork bifurcations.

1. **Introduction**

Localized excitations in arrays of coupled oscillators are often referred to as discrete breathers (DBs) or intrinsic localized modes (ILM) [1], [2]. The appearance of the DB solution is triggered by the discreteness of the system and nonlinearity of the evolution equations [2]. Such dynamical excitations normally exhibit exponential localization; in models with purely nonlinear inter-particle interaction stronger hyper-exponential localization is encountered [2]. These regimes have been observed and studied in various physical systems, and here we give several examples. Forced-damped coupled pendula and stability boundaries between different localized modes are studied in [3]. In [4], magnetic metamaterial breathers in energy conserving and dissipative systems are explored and coupling limit, stability and mobility properties are examined. Paper [5] demonstrates numerically and experimentally the excitation of the ILM as a result of subharmonic, spatially homogeneous driving. Several studies proved the existence of DBs in driven micromechanical silicon-nitride cantilever arrays [6-8]. In [9] it was examined how the activation of localized mode effects certain properties (thermal expansion, thermal and electrical conductivity) of metallic uranium and ionic sodium iodide crystals.

In many studies related to the DBs, conservative models are used [10]. In realistic systems the dissipation commonly cannot be neglected, and forcing of some kind should be applied to maintain the oscillations. When damping and forcing are not negligible, Hamiltonian approach is not applicable and stable localized excitation regimes become attractors of the dynamic flow. Mathematical analysis of such attractors in MDOF nonlinear systems is not a straightforward task. Consequently, most systems possessing the DBs are analysed by means of numerical simulations. Some recently studied models allow broader exploration of these dynamical regimes. In [11], [12] it was demonstrated that it is possible to obtain exact solutions for

DBs in damped chains with vibro–impact nonlinearity in the case of a homogeneous external forcing. Moreover, it was found possible to determine a zone of existence of the DB in a space of parameters and to reveal its stability thresholds.

Instead of direct external forcing, one can counter-balance the dissipation by internal generation of energy inside the system. Such phenomenon, referred to as self-excitation, is well-known in physics [13]. This phenomenon is usually realized through an internal feedback mechanism. Few of the many examples of self-oscillatory systems are the Huyghens pendulum clock [14], flutter of suspension bridges (e.g. Tacoma Narrows Bridge in 1940 [15]), the human voice [16] and biochemical processes [17], [18].

Such systems often take advantage of energy transform in order to produce self-sustaining dynamics. Dynamics of various nanomechanical systems can be described in terms of the self-excitation. Optomechanical systems can be driven by bolometric [19] or radiation pressure feedback [20]. In nanoelectromechanical systems (NEMS), internal feedback has been achieved by field emission of vibrating nanowires subject to a DC voltage [21], the periodic charging of a nanowire [22], transport through a carbon nanotube quantum dot mediated by the backaction of tunnelling single electrons [23]. Recently, a thermodynamic feedback of a piezoresistive resonator was employed to achieve self-oscillation [24].

A famous classical example of generation of self-oscillatory motion driven by conversion of electricity to mechanical oscillation is Franklin bell [25]. The system consists of two bells, one connected to a lightning rod attached to a chimney, the other one - to the ground. A metal ball hung between the two bells was warning Franklin that storm was coming by swinging repeatedly from one bell to the other as result of repetitive charge and discharge periods.

A particularly significant study of the self-excitation in nano-mechanical systems realized the idea of Franklin bell on nano-level (Gorelik [26]). In this system, the Coloumb blockade effect in combination with soft links attaching a metallic grain to electrodes was found to result in significant oscillations. At sufficiently large value of the bias voltage, the oscillation amplitude increased until the balance between energy dissipation and energy absorption was achieved, and the system reached the stable self-excitation regime. A cyclic change in direction is caused by the repeated pick up of N electrons at the source electrode and their transfer to the drain electrode. As a result, the sign of the net grain charge alternated leading to the oscillatory motion of the grain. In this way, the "electron shuttle" mechanism for the charge transport was realized.

Electron-shuttles were further investigated in several studies and demonstrated rather onterstng and beneficial properties [27-32]. One of the most interesting opportunities is a suggestion of extremely accurate discrete current source [27], [28]. In [29] the regions of self-excitations were derived analytically and verified experimentally. Bifurcations driven by change of system parameters such as electromechanical-coupling were demonstrated. Electron-shuttles were found to also operate at radio frequencies (RF), possess a set of resonance frequencies, and exhibit Coulomb blockade even at room temperature [30]. Spontaneous symmetry breaking has also been observed in these systems [31], revealing Arnold tongues in the frequency domain [32].

As a result of their diversity, electron-shuttles may possibly encourage progress in such fields as Metrology [33], Ferromagnetic island may encourage investigation of Kondo shuttling [34] and realization of spintronic devices (e.g. mechanical spin valves) [35], a superconducting shuttle may allow to target mechanically mediating phase coherence in a nanostructured device [36].

A self-excited system such as Franklin bell or the "electron-shuttle" may be mathematically represented as a single degree-of-freedom oscillator restricted by two inelastic constraints and subjected to constant force, which direction is reversed upon reaching the constraints. This work will consider dynamical behaviour of arrays of coupled self-excited oscillators of this type. Special attention will be paid to localization patterns and their stability. As the first problem to address, we consider the discrete breather, localized on a single excited oscillator. As a simple model, we adopt an infinite chain restricted by inelastic constraints, where the central particle is subjected to self-excitation of the specified type. The analysis presented in this paper focuses on derivation of analytic solutions for the discrete breathers, exploration of existence and stability boundaries in a parametric space and numerical verification of the results.

2. **Description of the model and analytical solutions**
   a. **Model description**

As it was mentioned before, the single self-excited model "Franklin bell" is a single particle which movement is restricted by two inelastic constraints. Interaction of the particle with the constraints is described by Newton impact model with restitution coefficient $k$. These constraints are, in turn, positively and negatively charged electrodes. The particle is connected to a fixed location through linear elastic leaf-spring with stiffness $\gamma$. We suggest for simplicity that after each re-charging the particle carries the same absolute amount of electric charge – positive or negative, depending on which electrode was impacted last. If the applied voltage is constant, one can adopt that between the impacts with the electrodes the particle is subject to action of force with constant magnitude $f$. Direction of the force is determined by the sign of the charge carried by the particle at the moment and is reversed after each impact. The sketch of the single-DOF basic model is shown in *Figure 1*. The distance between the electrodes is set to 2 without loss of generality. The stable self-excited regime of this single-DOF model is described by simple analytic solution; dependence of frequency of oscillations on the system parameters is expressed by Eq. (1).

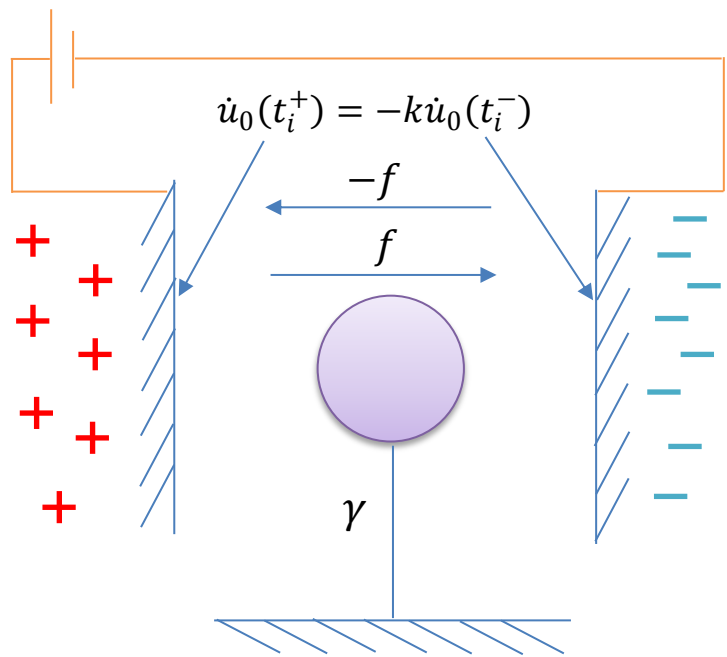

*Figure 1 - Sketch of the single particle system*

$$\omega_0 = \frac{\pi\sqrt{\gamma}}{\cos^{-1}\dfrac{(f+\gamma)k+(f-\gamma)}{(f+\gamma)+(f-\gamma)k}} \quad (1)$$

The model of coupled Franklin bell oscillators addressed in this work is a chain of particles coupled by linear elastic springs with unit stiffness (without loss of generality). Similarly to the single-DOF model of the Franklin bell outlined above, the movement of each particle is restricted by two inelastic electrodes with restitution coefficient $k$. To construct the simplest possible localized solution, we adopt that only the central particle ($n = 0$) impacts the constraints and, consequently, carries the charge. Similarly to the single-DOF system, this particle is subject to action of force with constant magnitude $f$, which direction is reversed upon reaching the constraints. A sketch of the system is shown in Figure 2.

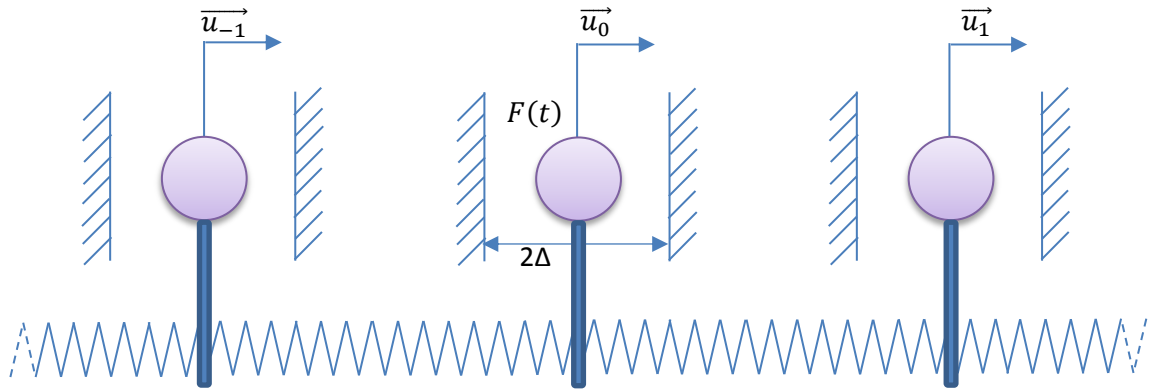

*Figure 2 - Sketch of the model system*

We suggest that the DB solution is periodic in time. In this case, the impacts with the electrodes can be expressed as periodic pulses, which act on the particle $n=0$. Consequently, equations of motion of the model can be cast in the following form:

$$\ddot{u}_n + (2u_n - u_{n-1} - u_{n+1}) = F(t)\delta_{n0}$$

$$F(t) = 2p\sum_{j=-\infty}^{\infty}\left\{\delta[t+Tj] - \delta\left[t+T\left(j+\frac{1}{2}\right)\right]\right\} + G(t), \quad G(t) = \begin{cases} f, & t \in \left[0+Tk, \dfrac{T}{2}+Tk\right) \\ -f, & t \in \left(\dfrac{T}{2}+Tk, T(k+1)\right) \end{cases}$$

(2)

In Eq. (2) $\delta_{n0}$ is the Kronecker symbol, $\delta(t)$ is Dirac delta function. $2p$ is yet unknown amount of momentum transferred to the central particle at each impact. Function $G(t)$ represents the alternating driving force, which is a rectangular wave with amplitude $f$ and period time $T$. The oscillation period $T$ is unknown and should be determined in the course of treatment.

It is important to notice that equation (2) assumes symmetric solution and periodic response. Besides, it is adopted that only particle $n=0$ reaches the constraints. Any solution that doesn't comply with these assumptions will require different formulation.

Without affecting the generality, in Eq. (2) the mass of each particle and clearance $\Delta$ (se Figure 2) are set to unity. One should supplement Eq. (2) by explicit formulation of the impact conditions. At the impact instance of the central particle $t_{i,o}$, the following conditions are satisfied:

$$t = t_{i,0}; \quad u_0(t_{i,o}) = \pm 1$$
$$\dot{u}_0(t_{i,0}^+) = -k\dot{u}_0(t_{i,0}^-) \tag{3}$$

In this model, the impact losses are the only source of damping in the system. The notations $\dot{u}_0(t_{i,0}^+)$, $\dot{u}_0(t_{i,0}^-)$ denote the velocities immediately after and before impacts, respectively. Naturally, the diapason of the restitution coefficient is $0 \leq k \leq 1$.

The forcing in (2) can be rewritten (in the sense of distributions) in terms of Fourier series with frequency $\omega_0 = 2\pi/T$:

$$F(t) = \frac{4}{\pi}\sum_{l=0}^{\infty}\left\{p\omega_0 \cos\left[\omega_0(2l+1)t\right] + \frac{f}{2l+1}\sin\left[\omega_0(2l+1)t\right]\right\} =$$
$$= \frac{4}{\pi}\sum_{l=0}^{\infty}\sqrt{p^2\omega_0^2 + \frac{f^2}{(2l+1)^2}}\sin\left[\omega_0(2l+1)t + \alpha_l\right]; \quad \alpha_l = \tan^{-1}\left(\frac{p\omega_0(2l+1)}{f}\right) \tag{4}$$

**b. Localized breather in the attenuation zone of the chain**

In this subsection we derive the analytical solution for the DB, similarly to the derivation in paper [11]. The periodic solution has the following form:

$$u_n = \sum_{l=0}^{\infty} u_{n,l} \sin\left[\omega_0(2l+1)t + \alpha_l\right] \tag{5}$$

Substitution of (4), (5) into (2) yields:

$$-\omega_0^2(2l+1)^2 u_{n,l} + (2u_{n,l} - u_{n-1,l} - u_{n+1,l}) = \frac{4}{\pi}\sqrt{p^2\omega_0^2 + \frac{f^2}{(2l+1)^2}}\delta_{n0} \tag{6}$$

For $n \neq 0$ Eq. (6) is homogenous. The solution for $n \to \pm\infty$ should vanish, therefore one obtains:

$$u_{n,l} = u_{0,l}\xi^{|n|}$$
$$\xi = -\frac{\left[\omega_0^2(2l+1)^2 - 2\right] + \sqrt{\omega_0^4(2l+1)^4 - 4\omega_0^2(2l+1)^2}}{2} \tag{7}$$

For $n = 0$ substitution of Eq. (7) into Eq. (6) yields then expression for the Fourier coefficients with $n = 0$:

$$u_{0,l} = -\frac{4}{\pi} \frac{\sqrt{p^2 \omega_0^2 + \frac{f^2}{(2l+1)^2}}}{\sqrt{\omega_0^4 (2l+1)^4 - 4\omega_0^2 (2l+1)^2}} \tag{8}$$

Therefore, by substituting Eq. (7), (8) into Eq. (5) the general expression for the periodic localized solution is obtained:

$$u_n = -\frac{4}{2^{-|n|}\pi} \times$$

$$\times \sum_{l=0}^{\infty} \sqrt{p^2 \omega_0^2 + \frac{f^2}{(2l+1)^2}} \frac{\left(2 - \omega_0^2 (2l+1)^2 - \sqrt{\omega_0^4 (2l+1)^4 - 4\omega_0^2 (2l+1)^2}\right)^{-|n|}}{\sqrt{\omega_0^4 (2l+1)^4 - 4\omega_0^2 (2l+1)^2}} \sin\left(\omega_0 (2l+1)t + \alpha_l\right) \tag{9}$$

The unknowns of expression (9) are $\omega_0$, $p$. These unknowns are found with the help of impact conditions (3). For the sake of simplicity, the first impact is set to occur at $t = 0$:

$$\begin{aligned} u_0(0) &= -1 \\ \dot{u}_0(0^+) &= -k\dot{u}_0(0^-) \end{aligned} \tag{10}$$

By satisfying condition $u_0(0) = -1$ one obtains:

$$p = \frac{\pi \omega_0}{4} \frac{1}{\sum_{l=0}^{\infty} \frac{1}{(2l+1)\sqrt{(2l+1)^2 - \frac{4}{\omega_0^2}}}} \tag{11}$$

Impact condition $\lim_{\varepsilon \to 0} \dot{u}_0(0+\varepsilon) = -\lim_{\varepsilon \to 0} k\dot{u}_0(0-\varepsilon)$ further yields:

$$p = \frac{4f}{\pi \omega_0} \frac{1+k}{1-k} \sum_{l=0}^{\infty} \frac{1}{(2l+1)\sqrt{(2l+1)^2 - \frac{4}{\omega_0^2}}} \tag{12}$$

Combining Eq. (11) and (12), one obtains expressions (one of them implicit) for the unknowns of expression (9):

$$p^2 = f \frac{1+k}{1-k} \tag{13}$$

$$\sum_{l=0}^{\infty}\frac{1}{(2l+1)\sqrt{\omega_0^2(2l+1)^2-4}}=\frac{\pi}{4\sqrt{f}}\sqrt{\frac{1-k}{1+k}} \qquad (14)$$

From Eq. (13) it can be seen that the momentum transferred to the main particle is a function of two system parameters: $f, k$, and can be derived explicitly. An important property of the momentum is that it is directly related to the forcing, which results in the possibility of extremely weak impacts for very small forcing.

The oscillation frequency $\omega_0$, cannot be explicitly extracted from Eq. (14). Though, it is seen that the frequency depends on system parameters ($f, k$). Two important conclusions can be drawn from Eq. (14).

   1) The solution can exist at all frequencies above $2$ (i.e. in the whole attenuation zone (AZ) of the linear chain with unit stiffness and unit mass).

   2) Approach of the frequency towards the AZ boundary corresponds to the vanishing forcing, i.e. $f \to 0 \Leftrightarrow \omega_0 \to 2$.

A plot of $\omega_0$ as a function of $f, k$ is given in Figure 3.

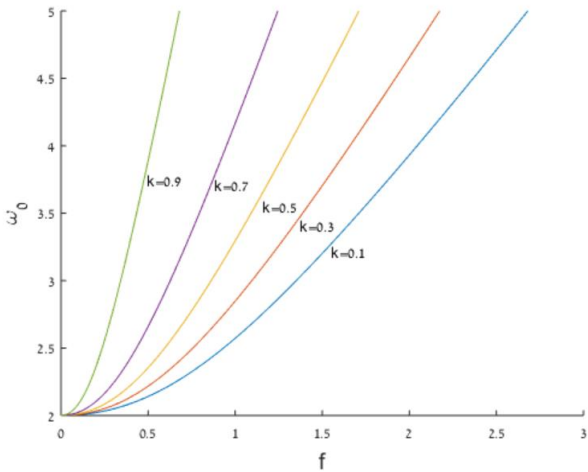

Figure 3 - Illustration of frequency of localized solution as a function of system parameters.

Once the periodic solution is found, it is natural to explore the extent of its existence. One can easily see that the maximal existence limit is the propagation zone (PZ) boundary $\omega_0 = 2$. In order to find the actual existence zone in the space of parameters and to ensure the self-consistency of the solution, it is necessary to satisfy the condition that the particles with $n \neq 0$ never reach the impact constraints. Thus, one should demand that for all particles $n > 0$:

$$\max_t \left[u_n(t)\right] \leq 1 \qquad (15)$$

In Appendix A it is demonstrated that condition (15) is satisfied for any set parameters as long as $\omega_0 > 2$. Thus, this analysis leads to conclusion that the existence zone of the periodic solution is as broad as the AZ of the chain. This, however, by no means guarantees the stability of this solution. The latter is analyzed in Section 3.

### c. Localized breather in the propagation zone of the chain

As will be further presented in the numerical section, at certain parameters, an additional localized solution in the PZ exists for this system, and it coexists with the DB solution derived above. To demonstrate this possibility, we construct below the solution, where only the first harmonics belongs to the PZ, while the rest infinite number of harmonics are situated in the AZ. The proposed solution is a superposition of the two essentially different forms.

First we present the system in an equivalent form which results from the symmetrical forces induced on the central particle by $u_1, u_{-1}$.

$$\begin{cases} \ddot{u}_0 + 2(u_0 - u_1) = \dfrac{4}{\pi} \sum_{l=0}^{\infty} \sqrt{p^2 \omega_0^2 + \dfrac{f^2}{(2l+1)^2}} \sin\left(\omega_0 (2l+1)t + \alpha_l\right) & n = 0 \\ \ddot{u}_n + (2u_n - u_{n-1} - u_{n+1}) = 0 & n > 0 \end{cases} \quad (16)$$

The suggested solution consists of two parts - irradiated (travelling) and exponentially localized:

$$u_n = \underbrace{-a\cos(\omega_0 t + \alpha_0 - kn)}_{travelling} + \underbrace{\sum_{l=1}^{\infty} u_{n,l} \sin\left[\omega_0(2l+1)t + \alpha_l\right]}_{localized} \quad (17)$$

By substituting the travelling part of solution (17) into (16) for $n=0$, one obtains the following equation:

$$\omega_0^2 a \cos(\omega_0 t + \alpha_0) - 2a\left[2\cos(\omega_0 t + \alpha_0)\sin^2\frac{k}{2} - \sin(\omega_0 t + \alpha_0)\sin k\right] = $$
$$= \frac{4}{\pi}\sqrt{p^2\omega_0^2 + f^2}\sin(\omega_0 t + \alpha_0) \quad (18)$$

From (18) it is easy to obtain the following expressions for wavenumber (it is, of course, common dispersion relation for the linear chain), and amplitude of the "propagating" harmonics :

$$k = 2\sin^{-1}\left(\frac{\omega_0}{2}\right) \quad (19)$$

$$a = \frac{2\sqrt{p^2\omega_0^2 + f^2}}{\pi \sin\left(2\sin^{-1}\left(\frac{\omega_0}{2}\right)\right)} \quad (20)$$

The travelling part in terms of $p, \omega_0$ is though:

$$u_{n-travelling} = \frac{2\sqrt{p^2\omega_0^2 + f^2}}{\pi \sin\left(2\sin^{-1}\left(\frac{\omega_0}{2}\right)\right)} \cos\left(\omega_0 t + \alpha_0 - 2n\sin^{-1}\left(\frac{\omega_0}{2}\right)\right) \quad (21)$$

The localized part of the solution has the form similar to that of the fully localized solution (9), without the first harmonics (summation starts at 1):

$$u_{n-localized} = -\frac{4}{\pi(2\gamma)^{-|n|}} \sum_{l=1}^{\infty} \sqrt{p^2\omega_0^2 + \frac{f^2}{(2l+1)^2}} \frac{\left(2-\omega_0^2(2l+1)^2 - \sqrt{\omega_0^4(2l+1)^4 - 4\omega_0^2(2l+1)^2}\right)^{-|n|}}{\sqrt{\omega_0^4(2l+1)^4 - 4\omega_0^2(2l+1)^2}} \sin(\omega_0(2l+1)t + \alpha_l) \quad (22)$$

The two unknowns are $p, \omega_0$. Impact conditions (3) are sufficient to determine them. The following expressions are obtained:

$$p = \frac{\pi\omega_0}{4}\left[1 - \frac{2f}{\pi \sin\left(2\sin^{-1}\left(\frac{\omega_0}{2}\right)\right)}\right] \frac{1}{\sum_{l=1}^{\infty} \frac{1}{(2l+1)\sqrt{(2l+1)^2 - \frac{4}{\omega_0^2}}}} \quad (23)$$

$$p = \frac{4f(1+k)}{\pi\omega_0} \frac{\sum_{l=1}^{\infty} \frac{1}{(2l+1)\sqrt{(2l+1)^2 - \frac{4}{\omega_0^2}}}}{\frac{2\omega_0^2(1+k)}{\pi}\sin\left(2\sin^{-1}\left(\frac{\omega_0}{2}\right)\right) + 1 - k} \quad (24)$$

By equating expressions (23), (24), the following transcendent equation for single unknown $\omega_0$ is obtained:

$$\frac{16f(1+k)}{\pi^2\omega_0^2} \frac{\left(\sum_{l=1}^{\infty} \frac{1}{(2l+1)\sqrt{(2l+1)^2 - \frac{4}{\omega_0^2}}}\right)^2}{\frac{2\omega_0^2(1+k)}{\pi}\sin\left(2\sin^{-1}\left(\frac{\omega_0}{2}\right)\right) + 1 - k} + \frac{2f}{\pi \sin\left(2\sin^{-1}\left(\frac{\omega_0}{2}\right)\right)} - 1 = 0 \quad (25)$$

Eq. (25) can be solved numerically, and one can directly obtain values of $p, a$ by substitution into (23) and (20).

The existence of this solution depends on satisfying requirement (15). It was verified numerically for a wide range of parameters that this solution is self-consistent.

### d. Other solutions

So far, two, essentially different, exact solutions were derived analytically (and will be verified numerically in Section 4). This, naturally, raises the question about other stable attractors in the system. Theoretically, a number of additional solutions can be derived for the suggested model, e.g. the solutions with different number of harmonics in the propagation zone. We restrict ourselves by two solutions presented so far for two reasons: 1) These solutions are structurally different while additional solutions are expected to have structure similar to the solution in the PZ; 2) Numerical analysis implies that the solutions presented above are in a sense the most typical, while the other solutions are expected to be stable in considerably narrower regions of system parameters.

### 3. Stability analysis
### a. Localized AZ solution

In this section a linear stability analysis of the exponentially localized breather in the attenuation zone is carried out by means of Floquet theory. The theory requires computation of the monodromy matrix which describes the evolution of perturbation of the state vector (26), for a single period. According to Floquet theory, the loss of stability occurs when at least one eigenvalue leaves the unit circle in the complex plane. In our case, each period of motion consists of 2 impacts and 2 periods of linear evolution between the impacts. This can be generally represented as product of 4 matrices (27), where $Q$ describes the evolution between impacts, and $S$ is the saltation matrix [37] which describes the evolution of the perturbed phase flow at the impact hypersurfaces:

$$v = \begin{bmatrix} u_{-n} & \cdots & u_{-1} & u_0 & u_1 & \cdots & u_n & \dot{u}_{-n} & \cdots & \dot{u}_{-1} & \dot{u}_0 & \dot{u}_1 & \cdots & \dot{u}_n \end{bmatrix}^T \tag{26}$$

$$M = QSQS, \quad Q = e^{\frac{\pi}{\omega_0}A} \tag{27}$$

The evolution of the small perturbation between successive impacts is given by block construction in (28), where $I$ is an $2n + 1 \times 2n + 1$ identity matrix, 0 is a $2n + 1 \times 2n + 1$ zero matrix, and Laplace adjacency matrix $L$ is given below in (29):

$$Q_{4n+2\times 4n+2} = \begin{pmatrix} 0 & I \\ L & 0 \end{pmatrix} \tag{28}$$

$$L_{2n+1\times 2n+1} = \begin{pmatrix} -1 & 1 & 0 & \cdots & 0 \\ 1 & -2 & \ddots & \ddots & \vdots \\ 0 & \ddots & \ddots & \ddots & 0 \\ \vdots & \ddots & \ddots & -2 & 1 \\ 0 & \cdots & 0 & 1 & -1 \end{pmatrix} \tag{29}$$

In current work both acceleration and velocity are discontinuous at impacts; this issue effects the structure of the saltation matrix. It is presented by block construction in (30), where 0 is a $2n + 1 \times 2n + 1$ zero matrix, $K, C$ are given in (31). The only particle which is subjected to impacts is the central one, and for that reason the only non-zero term in matrix $C$ and the only non unity term in the diagonal of matrix $K$ is $(n + 1, n + 1)$.

$$S_{4n+2 \times 4n+2} = \begin{pmatrix} K & 0 \\ C & K \end{pmatrix} \tag{30}$$

$$C_{2n+1\times 2n+1} = \begin{pmatrix} 0 & \cdots & 0 & 0 & 0 & \cdots & 0 \\ \vdots & \ddots & \vdots & \vdots & \vdots & \ddots & \vdots \\ 0 & \cdots & 0 & 0 & 0 & \cdots & 0 \\ 0 & \cdots & 0 & \dfrac{\ddot{u}^+ + k\ddot{u}^-}{\dot{u}^-} & 0 & \cdots & 0 \\ 0 & \cdots & 0 & 0 & 0 & \cdots & 0 \\ \vdots & \ddots & \vdots & \vdots & \vdots & \ddots & \vdots \\ 0 & \cdots & 0 & 0 & 0 & \cdots & 0 \end{pmatrix} \quad K_{2n+1\times 2n+1} = \begin{pmatrix} 1 & \cdots & 0 & 0 & 1 & \cdots & 0 \\ \vdots & \ddots & \vdots & \vdots & \vdots & \ddots & \vdots \\ 0 & \cdots & 1 & 0 & 0 & \cdots & 1 \\ 0 & \cdots & 0 & -k & 0 & \cdots & 0 \\ 1 & \cdots & 0 & 0 & 1 & \cdots & 0 \\ \vdots & \ddots & \vdots & \vdots & \vdots & \ddots & \vdots \\ 0 & \cdots & 1 & 0 & 0 & \cdots & 1 \end{pmatrix} \quad (31)$$

To explore the stability of the DB solution, we compute numerically the eigenvalues of the monodromy matrix. It is impossible to consider the infinite number of particles; therefore, the monodromy matrix is truncated at certain number *n*. It was found that the stability thresholds converge as the number of particles increases. In all stability calculations for the localized breather, $n = 300$ was found to be satisfactory. An example is given in *Figure 4*. The stability boundary in the $k - f$ plane is presented in *Figure 5*. The loss of stability at the boundary is realized through Neimark-Saker bifurcation. It is important to notice that close to the stability boundary, this solution coexists with the solution in the PZ; stability of the latter will be discussed in the next subsection. Therefore, in order to numerically verify the boundary of stability, it is necessary to select initial conditions very close to the desired solution. The issue of coexistence will be further addressed in subsection 3c and in Section 4.

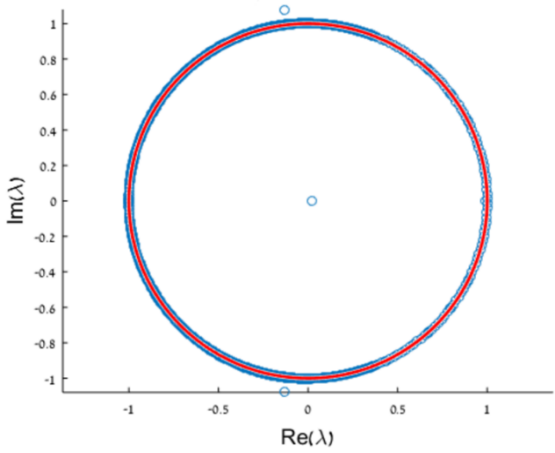

*Figure 4 - Floquet multipliers at the loss of stability of the localized solution by Neimark-Sacker bifurcation, system parameters: $k = 0.5, F = 0.33, n = 300$*

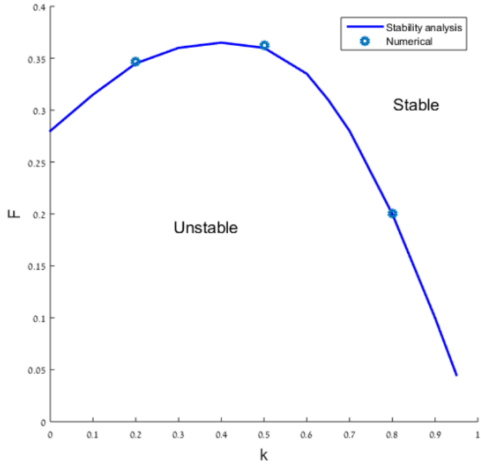

*Figure 5 –Stability boundary of the localized breather in the AZ; At forcing below the shown line, the stability is lost by Neimark-Sacker bifurcation, system parameters: $k = 0.5, n = 300$ (in stability analysis)*

### b. Stability of the PZ DB.

In the infinite chain, the propagating solution consists of a propagating wave and an exponentially localized component. Here, in order to perform the stability analysis, we substitute the initial infinite chain by a symmetric (with respect to the impacting particle) finite segment of the chain with oscillatory profile similar to the PZ DB (17-25). We also assume the boundary conditions at the edges of the selected segment that exactly reproduce the forces induced by the omitted parts of the chain. However, variation on these boundary conditions is equal to zero, so they have no effect on the monodromy matrix. Then, we can explore the stability of the PZ DB with respect to perturbations that are localized at the selected fragment of the infinite chain. Control of convergence of the stability boundary is achieved through exploration of longer fragments. The shape of the monodromy matrix is not different from the one derived for the AZ DB. Therefore, the only parameters which determine the Floquet multipliers are the DB frequency, and velocity and acceleration of particle $n=0$ before and after the impact. All of these parameters are calculated from expressions which were given in section 2c.

The solution in PZ was found to be stable between two limit curves in the $k-f$ plane. On the lower boundary, the solution loses stability through Neimark-Sacker bifurcation. On the upper boundary, the solution loses stability by pitchfork bifurcation for $0 < k < 0.45$ (example in *Figure 7*) and by Neimark-Sacker for $0.45 < k < 1$ (*Figure 6*). Again, due to coexistence of the solutions, in order to verify results numerically, initial conditions were chosen in close vicinity of the analytical solution.

The value of $n$ for simulations was selected such that the results were not influenced by further increasing of $n$ (the process converged). Here, on the lower boundary value of $n = 300$ was sufficient, while on the upper boundary it was required to use $n = 500$.

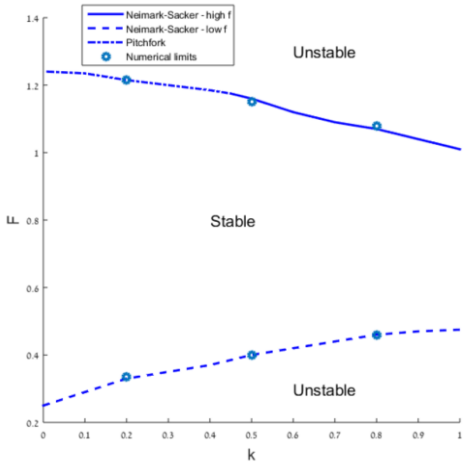

*Figure 6 –Stability boundary of the PZ DB. The bold line indicates the upper Neimark-Sacker boundary; The dashed line indicates the lower Neimark-Sacker boundary; The dash-dotted line indicates pitchfork bifurcation. System parameters: $k = 0.5$, upper bound: $n = 500$, lower bound: $n = 300$*

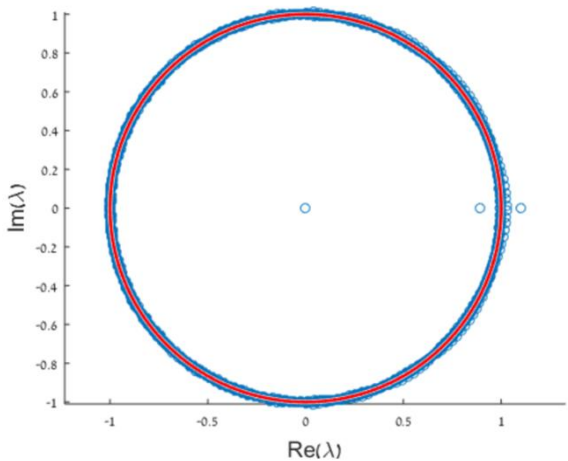

*Figure 7 - Floquet multipliers at the loss of stability of the propagating solution by pitchfork bifurcation, system parameters: $k = 0.2, F = 1.215, n = 500$*

### c. Overall observation of system stability thresholds

By comparing the stability zones for both solutions, it can be seen that they have overlapping zones of stability in the $k-f$ plane. These limits can be mapped to $k-\omega_0$ plane in order to examine the stability zones in terms of frequencies (*Figure 8*). The solid green lines indicate the existence limit of each solution: $\omega_0 = 2$ for the AZ DB, and $\omega_0 = 2/3$ for the PZ DB (at this frequency, the second harmonics enters the PZ). In both cases, the loss of stability close to these boundaries occurs by Neimark-Sacker mechanism. The frequency band of the PZ DB is considerably narrower than of the AZ DB. It is reasonable to assume that other solutions in propagation zone will be stable in narrow regions and will also lose stability, when their harmonics will approach the PZ boundary, by the generic Neimark-Sacker scenario. It is also worthy to mention that there are zones of frequencies where no one of the explored solutions is stable.

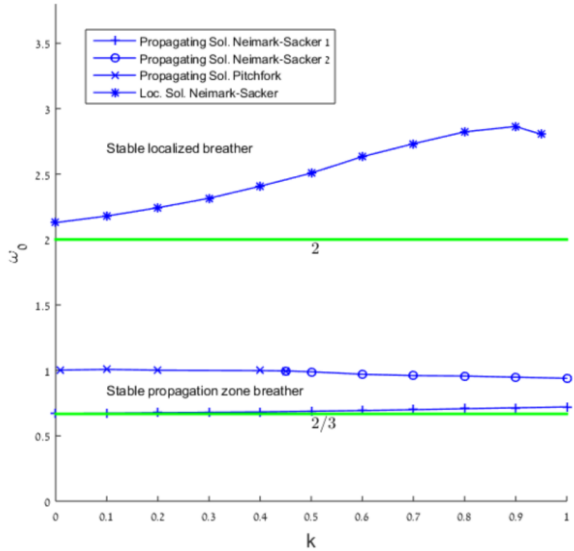

Figure 8 –Stability of localized and propagating solutions in plane of $k - \omega_0$.

### 4. Numerical Analysis

An example of the AZ DB obtained numerically, compared to the analytic solution, is presented in *Figure 9*. As one would expect from the exact solution, the analytical and numerical results match exactly. This solution is characterized by frequency in the AZ, and exponential decay of the amplitudes along the chain. In this example the amplitude of the nearest neighbour is $u_1 = 0.09$.

The second attractor described was the PZ. As was shown in the stability analysis, this solution co-exists with other solutions and approach to the vicinity of this attractor depends on careful choice of the initial conditions. An example of the solution is shown in *Figure 10*. The PZ DB solution has longer period and non-zero asymptotic value of the neighbours amplitude (in this example – 0.88). Another phenomenon illustrated by *Figure 9* and *Figure 10* is the coexistence in the system, as both solutions were obtained for the same system parameters with different initial conditions.

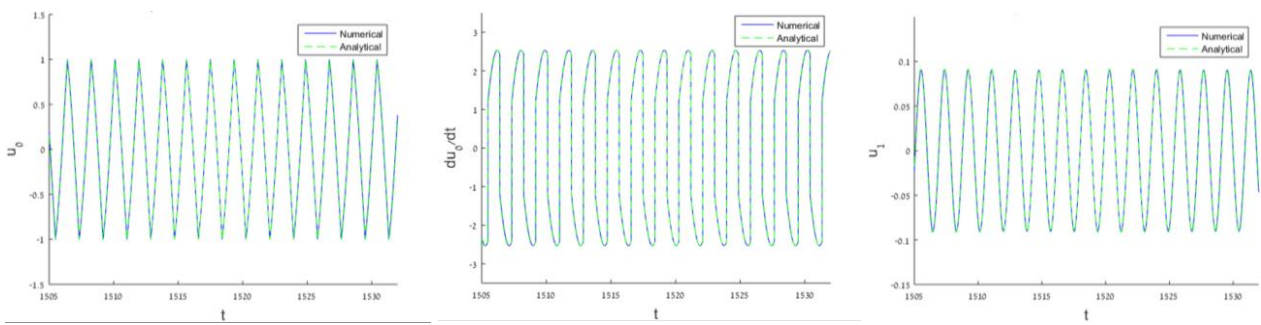

Figure 9 – Localized solution (numerical – blue, solid, analytical – green, dashed) $k = 0.5, F = 1.1, n = 1001, u_0(0) = 0.05, \dot{u}_0(0) = 3.0$

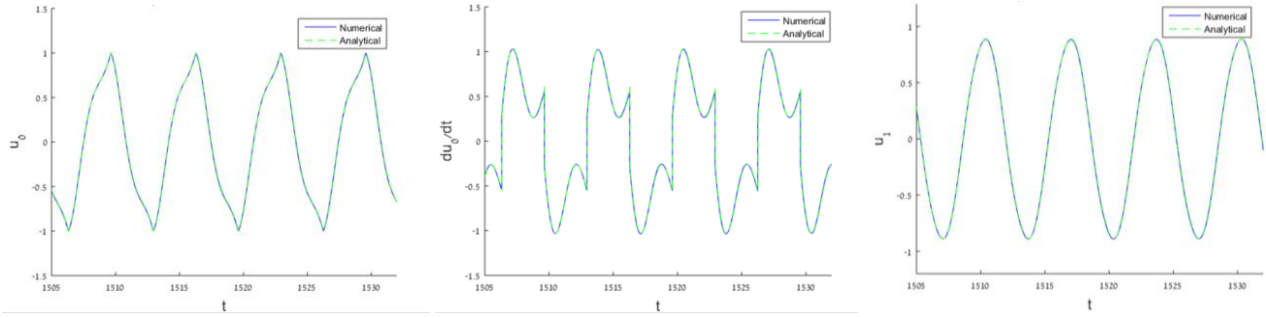

*Figure 10 – Propagating solution (numerical – blue, solid, analytical – green, dashed) $k = 0.5, F = 1.1, n = 1001, u_0(0) = 0.05, \dot{u}_0(0) = 0.1$*

Another way to examine coexistence in the system is by presenting both solutions in a certain projection of the state space. In *Figure 11* the two solutions are illustrated in a plane of $u_0 - \dot{u}_0$ with the particular initial condition that led to each one of the solutions (localized solution – 'o', propagating solution – '*'). The initial conditions for both solutions have 2 nonzero coordinates $u_0, \dot{u}_0$ (and can be represented in the 2D projection of *Figure 11*). Here, it is seen that stable attractors can be reached even if system is instantiated at a point distant from the analytic solution in the state space.

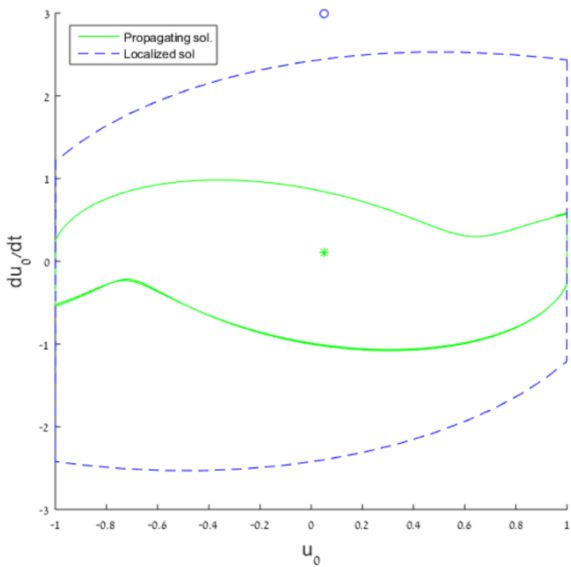

*Figure 11 – Coexistence of two solutions at same system parameters, dashed blue line - localized solution obtained for initial conditions $u_0(0) = 0.05, \dot{u}_0(0) = 3.0$, solid green line - propagating solution obtained for initial conditions $u_0(0) = 0.05, \dot{u}_0(0) = 0.1$, System parameters: $k = 0.5, F = 1.1, n = 501$*

It was shown that the localized solution loses its stability through Neimark-Sacker bifurcation. In order to explore the state of the system after the loss of stability, the forcing parameter was gradually decreased to the precise limit at which a modulated response, characteristic to Neimark-Sacker bifurcation, was achieved. The modulated response is most easily noticed in the response of the first neighbours. $(u_1, u_{-1})$, an example is shown in *Figure 12*. The FFT of this response provides the frequency of modulation, which in the case of the parameters in *Figure 12* is $\omega_{modulation} = 1.81 \left[\frac{rad}{sec}\right]$.

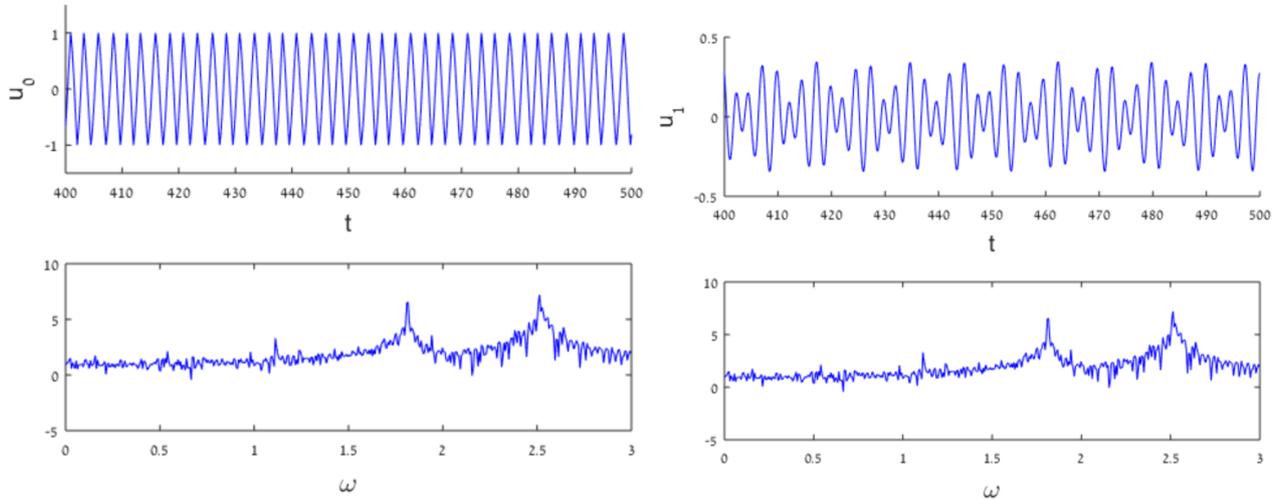

*Figure 12 – Response at the Neimark-Sacker bifurcation, parameters:* $\gamma = 0.1, k = 0.5, F = 0.38753114, n = 1001, u_0(0) = 0.05, \dot{u}_0(0) = 3.0$

In the stability analysis of the PZ DB it was found that the solution has a region where it loses stability through the pitchfork bifurcation. In *Figure 13* a typical numerical response is presented at the pitchfork boundary.

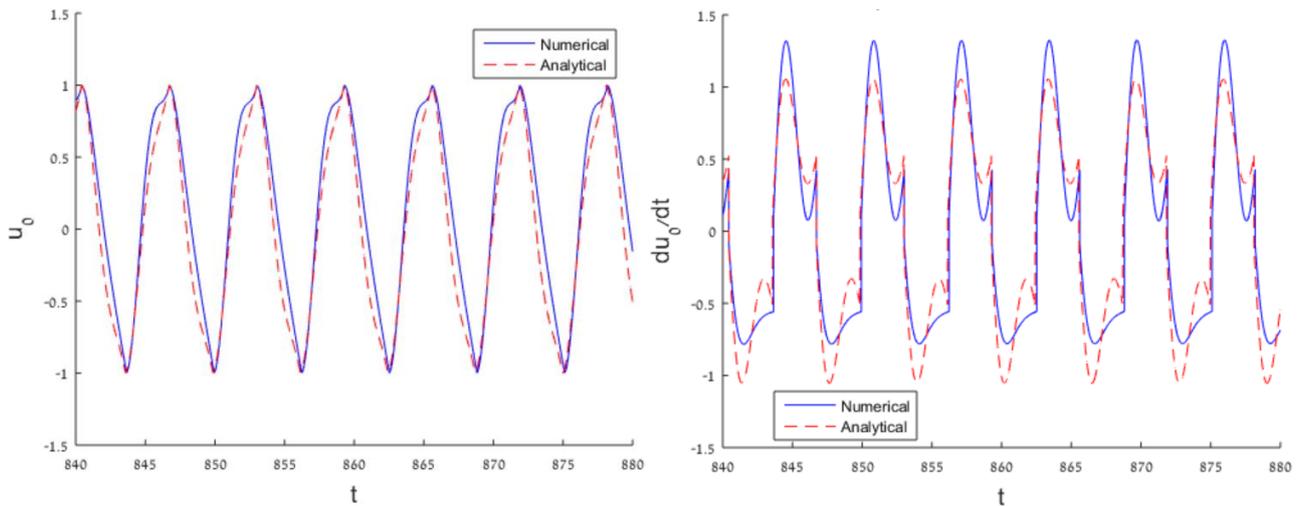

*Figure 13 - Loss of stability of the propagating solution by the pitchfork bifurcation, parameters:* $k = 0.2, F = 1.215, n = 2001, initial\ conditions - on\ analytical\ solution\ for\ the\ symmetric\ breather$

5. **Concluding remarks**

Current work addressed an infinite symmetrical chain of coupled oscillators, subject to the self-excitation. Analytical solutions for the DBs were derived for two qualitatively different dynamical regimes. The first solution was derived for the DB in the attenuation zone of the chain and is characterized by exponential decay of the amplitude to zero. The second solution has the main frequency in the PZ of the chain; the other harmonics belong to the AZ and are exponentially localized. In this regime, the energy input due to the self-excitation is balanced not only by the inelastic impacts, but also by the energy irradiation into the chain.

The loss of stability for both solutions in the vicinity of the propagation zone occurred through the Neimark-Sacker bifurcation; it is reasonable to assume that the Neimark–Sacker bifurcation is a generic scenario that is triggered by interaction with the boundary of the propagation zone. Besides, the PZ breather has another stability boundary, where both Neimark-Saker and pitchfork bifurcations are encountered. It seems reasonable to suggest that this stability boundary is not generic and appears due to particularities of the forcing mechanism.

Regime of the self-excitation in the single-DOF oscillator (Figure 1) is stable for all combinations of parameters. Expansion of the model to a system of coupled self-excited oscillators brings about the nontrivial stability patterns. Therefore one can control the charge transport in the system of coupled oscillators by modification of the parameters. This possibility outlines possible applications for the localized solutions revealed in the paper.

The authors are grateful to Israel Science Foundation (grant 838/13) for financial support.

**Appendix A**

In this Appendix a proof of analytic solution self-consistency is presented. We demand that for all particles the following condition is satisfied:

$$\max_t \left[ u_n(t) \right] \leq 1 \qquad (A1)$$

Since the particles of the chain reach maximal amplitude in the vicinity of the propagation zone ($\omega_0 = 2$), in this Appendix the proof is narrowed to this region in the space of parameters.

Solution (9) can be rewritten to following form:

$$u_n = A_n(t) + R_n(t)$$

$$A_n = \frac{4(2)^{|n|}}{\pi} \sqrt{p^2 \omega_0^2 + f^2} \frac{1}{\left( \omega_0^2 - 2 + \sqrt{\omega_0^4 - 4\omega_0^2} \right)^{|n|} \sqrt{\omega_0^4 - 4\omega_0^2}} \sin(\omega_0 t + \alpha)$$

$$R_n = \frac{4 \cdot 2^n}{\pi} \sum_{l=1}^{\infty} \left[ \frac{\sqrt{p^2 \omega_0^2 + \frac{f^2}{(2l+1)^2}}}{\left( \omega_0^2 (2l+1)^2 - 2 + \omega_0 (2l+1) \sqrt{\omega_0^2 (2l+1)^2 - 4} \right)^{|n|} \sqrt{\omega_0^4 (2l+1)^4 - 4\omega_0^2 (2l+1)^2}} \cdot \sin(\omega_0 (2l+1)t + \alpha_l) \right] \qquad (A2)$$

In this expression, the exact solution is decomposed into two components - $A_n$ is the component with $\omega_0$ frequency, $R_n$ is the component with all other harmonics.

From (14) it is seen that:

$$f \to 0 \Leftrightarrow \omega_0 \to 2 \qquad (A3)$$

It is important to notice that these two limits occur simultaneously. Moreover, these limits are correct in the region of interest, where the exact solution reaches its maximal value. Therefore, it is sufficient to prove that $\max[u_n(t)] \leq 1$ for condition (A3).

Further, the following notations will be used:

$$\hat{A}_n = \max[A_n] \qquad \hat{R}_n = \max[R_n]$$

For $\hat{A}_n$ the following estimation can be derived:

$$\hat{A}_n \leq \frac{4(2)^{|n|}}{\pi \omega_0} \sqrt{p^2 \omega_0^2 + f^2} \frac{1}{\left(\omega_0^2 - 2 + \sqrt{\omega_0^4 - 4\omega_0^2}\right)^{|n|} \sqrt{\omega_0^2 - 4}} \tag{A4}$$

By using the fact that $\omega_0 > 2$, and the Riemann zeta function one can obtain the following estimation for $\hat{R}_n$.

$$\hat{R}_n < \frac{4 \cdot 2^n}{\pi} \sum_{l=1}^{\infty} \left[ \frac{\sqrt{p^2 \omega_0^2 + \frac{f^2}{(2l+1)^2}}}{\left(\omega_0^2(2l+1)^2 - 2 + \omega_0(2l+1)\sqrt{\omega_0^2(2l+1)^2 - 4}\right)^{|n|} \sqrt{\omega_0^4(2l+1)^4 - 4\omega_0^2(2l+1)^2}} \right] <$$

$$< \frac{2}{\pi} \sum_{l=1}^{\infty} \sqrt{p^2 \omega_0^2 + \frac{f^2}{(2l+1)^2}} \frac{1}{\omega_0^2 \omega_0^{2n}(2l+1)\left(2l^2 + 2l + (2l+1)\sqrt{l^2+l}\right)^{|n|} \sqrt{l^2+l}} < \tag{A5}$$

$$< \frac{1}{4^n \pi \omega_0^{2+2n}} \sum_{l=1}^{\infty} \sqrt{p^2 \omega_0^2 + \frac{f^2}{(2l+1)^2}} \frac{1}{l^{2+2n}} <$$

$$< \frac{1}{16^n \pi \omega_0^2} \left(\frac{4}{\omega_0^2}\right)^n \left(p \omega_0 \zeta(2+2n) + \frac{f}{2} \zeta(3+2n)\right)$$

By substitution of (13), Eq. (A4) can be rewritten the following form:

$$\hat{A}_n = \frac{4}{\pi \omega_0} \sqrt{f \omega_0^2 \frac{1+k}{1-k} + f^2} \frac{1}{\left(\frac{\omega_0^2 - 4\gamma}{2} + 1 + \frac{\omega_0}{2}\sqrt{\omega_0^2 - 4}\right)^{|n|} \sqrt{\omega_0^2 - 4}} \tag{A6}$$

By looking for the limit of (A6) for (A3) condition, one obtains:

$$\hat{A}_n \left(\begin{array}{c} \omega_0 \to 2 \\ f \to 0 \end{array}\right) = \frac{4}{\pi \omega_0} \sqrt{f \omega_0^2 \frac{1+k}{1-k} + f^2}_{\to 0} \frac{1}{\left(\underbrace{\frac{\omega_0^2 - 4}{2}}_{\to 0} + 1 + \frac{\omega_0}{2}\sqrt{\omega_0^2 - 4}\right)^{|n|} \sqrt{\omega_0^2 - 4}} =$$

$$= \frac{4}{\pi} \sqrt{f \frac{1+k}{1-k}} \frac{1}{\left(1 + \frac{\omega_0}{2}\sqrt{\omega_0^2 - 4}\right)^{|n|} \sqrt{\omega_0^2 - 4}} \tag{A7}$$

By substitution of (13) into (A7) one obtains the following approximation:

$$\hat{A}_n \begin{pmatrix} \omega_0 \to 2 \\ f \to 0 \end{pmatrix} = \frac{1}{\sum_{l=0}^{\infty} \frac{1}{(2l+1)\sqrt{\omega_0^2(2l+1)^2 - 4}}} \frac{1}{\sqrt{\omega_0^2 - 4}} \frac{1}{\left(1 + \frac{\omega_0}{2}\sqrt{\omega_0^2 - 4}\right)^{|n|}} =$$

$$= \frac{1}{1 + \sqrt{\omega_0^2 - 4}\underbrace{\sum_{l=1}^{\infty} \frac{1}{(2l+1)\sqrt{\omega_0^2(2l+1)^2 - 4}}}_{\to 0}} \frac{1}{\left(1 + \underbrace{\frac{\omega_0}{2}\sqrt{\omega_0^2 - 4}}_{\to 0}\right)^{|n|}} \approx \quad (A8)$$

$$\approx \left(1 - \sqrt{\omega_0^2 - 4}\sum_{l=1}^{\infty} \frac{1}{(2l+1)\sqrt{\omega_0^2(2l+1)^2 - 4}}\right)\left(1 - n\frac{\omega_0}{2}\sqrt{\omega_0^2 - 4}\right) =$$

$$\approx 1 - \sqrt{\omega_0^2 - 4}\left(n\frac{\omega_0}{2} + \sum_{l=1}^{\infty} \frac{1}{(2l+1)\sqrt{\omega_0^2(2l+1)^2 - 4}}\right)$$

From expressions (A5) and (A8), it can be seen that the consequent condition that has to be satisfied in order to ensure (15) is:

$$\frac{1}{16^n \pi \omega_0^2}\left(\frac{4}{\omega_0^2}\right)^n \left(p\omega_0\zeta(2+2n) + \frac{f}{2}\zeta(3+2n)\right) < \sqrt{\omega_0^2 - 4}\left(n\frac{\omega_0}{2} + \sum_{l=1}^{\infty} \frac{1}{(2l+1)\sqrt{\omega_0^2(2l+1)^2 - 4}}\right) \quad (A9)$$

Since $\sum_{l=1}^{\infty} \frac{1}{(2l+1)\sqrt{\omega_0^2(2l+1)^2 - 4\gamma}} > 0$, it is sufficient to demand:

$$\frac{1}{16^n \pi \omega_0^2}\left(\frac{4}{\omega_0^2}\right)^n \left(p\omega_0\zeta(2+2n) + \frac{f}{2}\zeta(3+2n)\right) < n\sqrt{\omega_0^2 - 4}\frac{\omega_0}{2} \quad (A10)$$

It is easy to see that if (A10) is satisfied for n=1, it is consequently satisfied for all n>1.

Substitution of n=1 into (A10) it can be rewritten to the form:

$$p\zeta(4) + \frac{f}{2\omega_0}\zeta(5) < 32\pi\sqrt{\omega_0^2 - 4}\left(\frac{\omega_0^2}{4}\right)^2 \quad (A11)$$

By using the fact that $\omega_0 > 2$ and substitution of (13) and (14), Eq. (A11) can be rewritten the following way:

$$\frac{\pi\zeta(4)}{4\sum_{l=0}^{\infty}\frac{1}{(2l+1)\sqrt{\omega_0^2(2l+1)^2-4}}}+\frac{\pi^2\frac{1-k}{1+k}\zeta(5)}{32\omega_0\left(\sum_{l=0}^{\infty}\frac{1}{(2l+1)\sqrt{\omega_0^2(2l+1)^2-4}}\right)^2}<32\pi\sqrt{\omega_0^2-4} \quad (A12)$$

Since $\sum_{l=1}^{\infty}\frac{1}{(2l+1)\sqrt{\omega_0^2(2l+1)^2-4}}>0$, it is sufficient to demand:

$$\frac{\pi\zeta(4)}{4\frac{1}{\sqrt{\omega_0^2-4}}}+\frac{\pi^2\zeta(5)}{32\omega_0\frac{1}{\omega_0^2-4}}<32\pi\sqrt{\omega_0^2-4} \quad (A13)$$

By rearranging Eq. (A13) one obtains:

$$\frac{\pi^4}{360}+\frac{\pi\zeta(5)}{32\omega_0}\sqrt{\omega_0^2-4}<32 \quad (A14)$$

By recalling the initial condition of this analysis ($\omega_0 \to 2$), one can see that (A14) is satisfied, for all parameters in the region of interest. Therefore, condition (A2) is satisfied, and the exact solution exists for any parameter in the attenuation zone of the chain.